\documentclass[aps,prd,twocolumn,showpacs,superscriptaddress]{revtex4-1}

\usepackage{graphicx}
\usepackage{dcolumn}
\usepackage{bm}
\usepackage{xcolor}

\usepackage{times}
\usepackage[normalem]{ulem}
\usepackage{units}
\usepackage[hypertexnames=false,linktocpage=true,colorlinks=true,linkcolor=blue,anchorcolor=blue,citecolor=blue,filecolor=blue,urlcolor=blue,bookmarksnumbered=true,pdfview=FitB,breaklinks=true]{hyperref}
\usepackage[normalem]{ulem}

\begin{document}

\title{Synthesis, structural and magnetic characterizations of Li$_4$Cu$_{1-x}$Ni$_x$TeO$_6$ ( $x$ = 0,  0.1, 0.2, 0.5, and 1)}
	\author{Ashiwini Balodhi}
	\affiliation{Department of Physics, University of Wisconsin-Milwaukee, Milwaukee, WI, 53211, USA.}
	\author{Brianna Billingsley}
	\affiliation{Department of Physics, University of Arizona, Tucson, AZ, 85721, USA}
	
	\author{Tai Kong}
	\affiliation{Department of Physics, University of Arizona, Tucson, AZ, 85721, USA}
	\author{Min Gyu Kim}
	\affiliation{Department of Physics, University of Wisconsin-Milwaukee, Milwaukee, WI, 53211, USA.}

  \email{Author to whom correspondence should be addressed: balodhia@uwm.edu, mgkim@uwm.edu}

\begin{abstract}

We investigated the effect of Ni doping in a recently proposed quantum spin liquid (QSL) candidate Li$_4$CuTeO$_6$. We performed a comprehensive study on the structural and magnetic properties. We find that the anti-site disorder between Li$^+$ and Cu$^{2+}$ persists until 50\% Ni doping in which Ni and Cu occupy different crystallographic sites. As a result, while Cu sits in both triangular and honeycomb layers in Li$_4$CuTeO$_6$,  Ni forms only honeycomb layer in Li$_4$NiTeO$_6$ and Li$_4$Cu$_{0.5}$Ni$_{0.5}$TeO$_6$. Our magnetic susceptibility measurements show that the Weiss temperature decreases from -145.68 K in Li$_4$CuTeO$_6$ to -6.15 K in Li$_4$NiTeO$_6$ as Ni doping increases, and find no hint of magnetic ordering or freezing down to 1.8 K. Our analysis implies the existence of abundant low energy excitations in these materials.

\end{abstract}

\maketitle

\section{Introduction} 
\setcounter{section}{1}
Triangular-lattice antiferromagnets host various exotic quantum states.~\cite{Anderson, Shimizu, Moessner1, Serbyn, Balents} One of the most interesting emergent quantum states is the quantum spin liquid (QSL) state that is claimed but still in debate, for example, in $\kappa$-(BEDT-TTF)$_2$Cu$_2$(CN)$_3$,~\cite{Kanoda, Matsuda} and EtMe$_3$Sb[Pd(dmit)$_2$]$_2$,~\cite{Matsuda1,Kato} 1-TaS$_2$,~\cite{Arcon} YbMgGaO$_4$,~\cite{Zhao, Zhang, Mourigal1} NaYbS$_2$, ~\cite{Baenitz1, Liu} and NaYbO$_2$. ~\cite{Walker, Nuttall1} Quantum fluctuations, topological order, and/or spin entanglement play an essential role in the exotic properties of the QSLs. While the search for the QSL state often focuses on identifying the absence of magnetic order at low temperatures where magnetic order is expected based on our conventional knowledge, confirming the QSL state is extremely challenging because states similar to QSL  can occur due to, for instance, Mermin-Wagner physics, \cite{Mermin, Cui1, Knolle} anisotropic spin interactions, the presence of disorder like impurities, and stacking faults.~\cite{Zhang1, Chernyshev, Okamoto, Syzranov, Katukuri, Smaha, Lee, Kimchi, Kundu}

\begin{figure}[b!]
   \centering
 \includegraphics[width=1\linewidth]{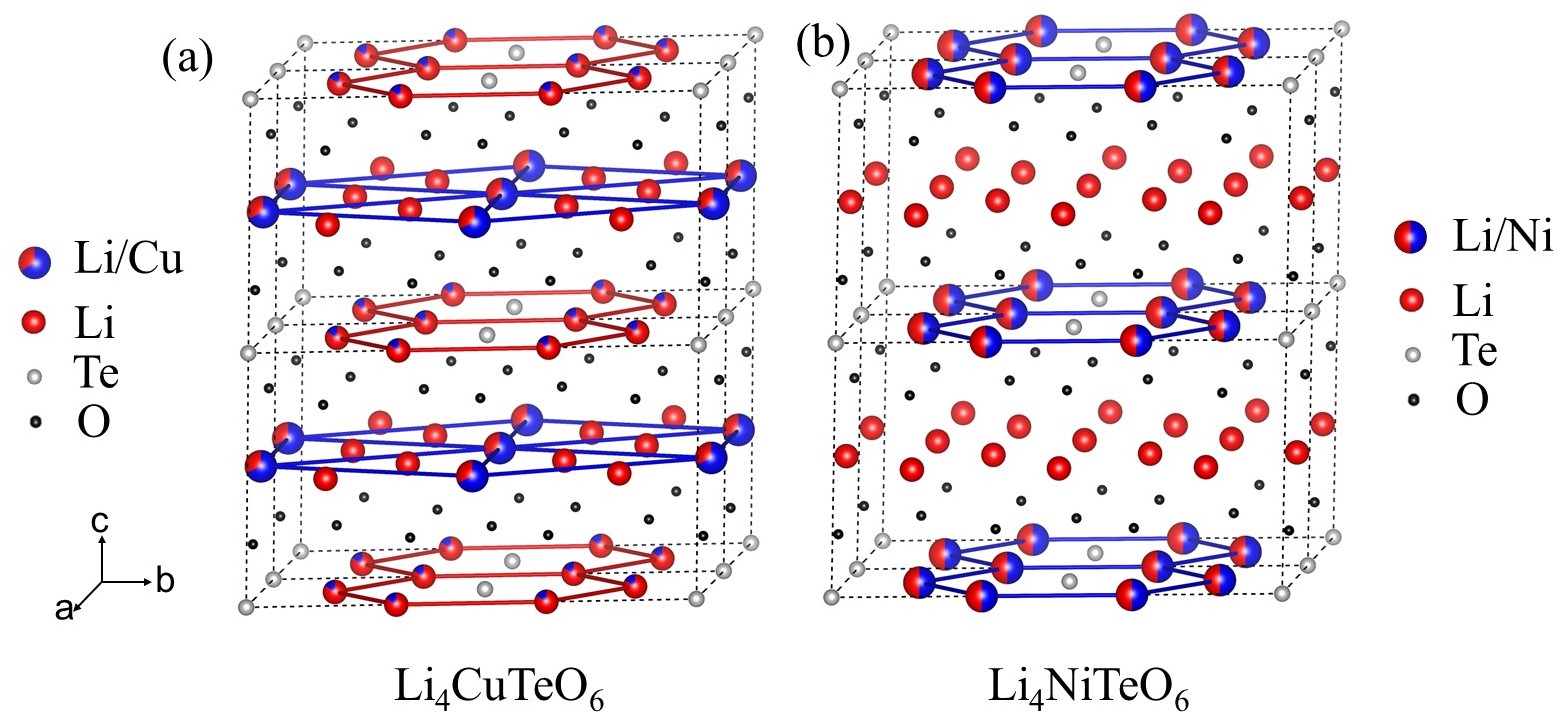}

   \caption{ (a) Crystal structures of Li$_4$CuTeO$_6$ and (b) Li$_4$NiTeO$_6$. Solid lines are honeycomb and triangular networks of Cu$^{2+}$ in Li$_4$CuTeO$_6$ and honeycomb network of Ni$^{2+}$ in Li$_4$NiTeO$_6$. Symbols with mixed colors (blue and red) indicate site mixing between Li and Cu or Li and Ni.}
  \label{unitcell}
\end{figure}

Recently, a Cu$^{2+}$ ($S$ = 1/2) based Li$_4$CuTeO$_6$ was shown to be a good QSL candidate.\cite{Khuntia} Li$_4$CuTeO$_6$  crystallizes in a monoclinic $\textit{C2/m}$ structure (space group No. 12) with $a$ $\approx$ 5.28 \AA, $b$ $\approx$ 8.82 \AA, $c$ $\approx$ 5.26 \AA and $\alpha$ = $\gamma$ = 90$^{\circ}$, $\beta \approx$ 113.17$^{\circ}$. \cite{Khuntia} An important structural feature of Li$_4$CuTeO$_6$ is that the triangular layers composed of Li$^+$ and Cu$^{2+}$ ions are stacked along the \textit{c} axis separated by honeycomb layers composed of Te$^{6+}$, Li$^+$, and Cu$^{2+}$.\cite{Khuntia, Uma} Interestingly, Li$^+$ and Cu$^{2+}$ ions share their Wyckoff sites with each other by partial occupancy in both the triangular (2$d$ site) and honeycomb layers (4$g$ site) as shown with mixed colors of atoms in Fig.~\ref{unitcell} (a). A QSL-like state is claimed to occur primarily due to the bond randomness because of the anti-site disorder between Li$^+$ and Cu$^{2+}$ ions on triangular layers.\cite{Khuntia} Also, Li$_4$CuTeO$_6$ does not show any magnetic order down to 45 mK.~\cite{Khuntia} Further, the strong antiferromagnetic (AFM) interaction is expected based on the Curie-Weiss temperature, $\theta_{\mathrm{CW}}$ $\approx$ -154  K.~\cite{Khuntia} Muon spin resonance measurements also confirm the absence of static local magnetic order down to 1.55  K.~\cite{Khuntia} It has been shown that the scaling behaviors as a function of $\mu_0 H/T$, $MT^{-0.15}$ in the magnetization and $T^{-0.5}$ in heat capacity, imply the presence of disorder-driven random spin-singlet ground state similar to the QSL state. ~\cite{Khuntia,Jose, Lee}

On the other hand, while Li$_4$NiTeO$_6$ crystallizes in the same monoclinic $C2/m$ structure,\cite{Tarascon, Bernd, Uma} previous studies show that Ni$^{2+}$ ions may go into the honeycomb layer and partially occupy 4$g$ site with Li$^+$ ions, but not in the triangular layers as shown in Fig.~\ref{unitcell} (b).\cite{Bernd} Therefore, the bond randomness that stems from partially occupied Cu$^{2+}$ ions in the triangular layer does not exist in Li$_4$NiTeO$_6$. However, it has been shown that Li$_4$NiTeO$_6$ may possess interesting quantum ground-state including QSL.~\cite{Mourigal} Magnetization and Electron Spin Resonance (ESR) studies of Li$_4$NiTeO$_6$ do not show any long-range magnetic order down to 1.8 K.~\cite{Bernd} Further, the Curie-Weiss temperature $\theta_{\mathrm{CW}} \approx$ -11  K and the low-temperature critical exponent ($\beta$ = 1.59) in ESR absorption line supports the quasi two-dimensional antiferromagnetism.~\cite{Bernd} A broad peak-like feature in the specific heat is observed at around $T = 0.3$  K, and shifts towards higher temperatures ($T \approx$ 2 K) under an applied magnetic field up to 4 T.~\cite{Mourigal} This observation also supports the quantum magnetic ground state in Li$_4$NiTeO$_6$.

Structurally, Li$_4$CuTeO$_6$ and Li$_4$NiTeO$_6$ form the same crystal structure, and site-mixing between Li and Cu/Ni happens in both compounds. The difference comes from specific crystallographic sites Cu or Ni occupies: Cu partially occupies sites in both honeycomb and triangular layers (4$g$ and 2$d$ sites, respectively) whereas Ni partially occupies the 4$g$ site in the honeycomb layer. It implies that their quantum ground states might be deeply related to the details of their crystal structures. Therefore it is essential to understand their structure changes with varying Cu/Ni contents and corresponding physical properties. Here, we investigate a systematic study of crystal structures and magnetic properties in Li$_4$Cu$_{1-x}$Ni$_x$TeO$_6$ with $x$ = 0, 0.1, 0.2, 0.5, and 1. We find that, with increasing Ni doping, the lattice parameters $a$ and $c$ decrease while the lattice parameter $b$ increases, leading to a unit cell volume reduction of  1\% for $x$ = 1. Interestingly, we observe that Ni$^{2+}$ partially occupy the 4$g$ site in the honeycomb layer and do not sit in the triangular layers in any Ni doping levels. We do not find any hint of magnetic ordering or freezing in all our samples down to 1.8 K, implying possible quantum magnetic ground states across all doping levels. An inverse power law behavior in the magnetic susceptibility suggests the abundance of low-energy excitations in these compounds.

\section{Experimental details}

 We synthesized polycrystalline Li$_4$Cu$_{1-x}$Ni$_x$TeO$_6$ ($x$ = 0, 0.1, 0.2, 0.5, and 1) compounds using the solid-state reaction technique. The starting materials Li$_2$CO$_3$ (99.999\% Alfa Aesar), TeO$_2$ (99.998\% Alfa Aesar), NiO (99.995\% Alfa Aesar), and CuO (99.995\% Alfa Aesar) were mixed in a stoichiometric ratio and placed in an alumina crucible with a lid and heated to 750 $^{\circ}$C in 4 h and held there for 18 h in air, after which furnace was cooled to room temperature. The resulting mixture was ground and pelletized, placed in an alumina crucible heated to 850 $^{\circ}$C in 4 h, and held there for 16 h before cooling to room temperature.

 The X-ray powder patterns (XRD) were obtained at room temperature using a Proto AXRD Benchtop Powder X-ray Diffractometer with Cu-K$\alpha$ radiation and were quantitatively analyzed using the Rietveld method by the GSAS software package.\cite{GSAS} The magnetic property measurements were performed down to 1.8 K using the VSM function in a Quantum Design Physical Property Measurement System Dynacool.

\section{Results}

\subsection{X-ray powder diffraction}

\begin{figure}[t!]
   \centering
 \includegraphics[width=1\linewidth]{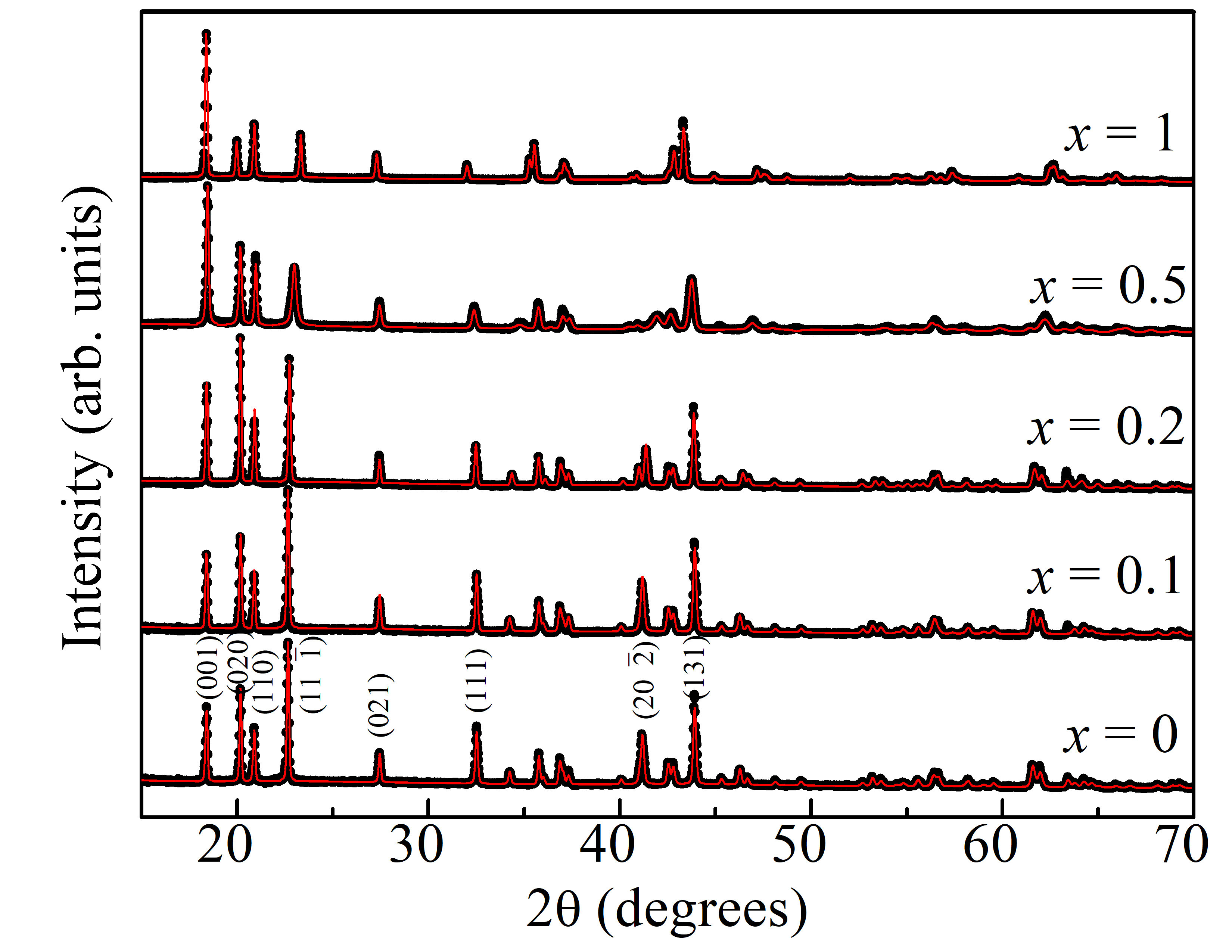} 
    \caption{ (Color online) Powder X-ray diffraction patterns for Li$_4$Cu$_{1-x}$Ni$_x$TeO$_6$ for $x$ = 0, 0.1, 0.2, 0.5, and 1 (black circle). The solid curve (red) through the data is the Rietveld refinement.}
  \label{PXRD}
\end{figure}

Figure \ref{PXRD} shows the room temperature XRD patterns of polycrystalline Li$_4$Cu$_{1-x}$Ni$_x$TeO$_6$ ($x$ = 0, 0.1, 0.2, 0.5, and 1) with the Rietveld refinement results. The XRD patterns were checked with the recently reported structure models for Li$_4$CuTeO$_6$ \cite{Uma} and Li$_4$NiTeO$_6$.\cite{Tarascon, Uma, Bernd} For the undoped  Li$_4$CuTeO$_6$ compound ($x =$ 0), our Rietveld refinement matches well with the previous report, yielding the unit cell parameters $a$ = 5.2692 (1) \AA, $b$ = 8.7887 (2) \AA, and $c$ = 5.2298 (1) $\AA$ with  $\alpha$ = $\gamma$ = 90$^{\circ}$, and $\beta$ = 113.043(7)$^{\circ}$. Importantly, as shown in Fig.~\ref{unitcell} and Table~\ref{table}, we find that a substantial site mixing is required between Li and Cu ions within not only the triangular layers but also the honeycomb layers , which is also consistent with previous reports.\cite{Khuntia, Uma}

 \begin{table*}[!]
\centering
\caption{Structural parameters obtained from a Rietveld refinement of room temperature powder X-ray patterns for Li$_4$Cu$_{1-x}$Ni$_x$TeO$_6$ ($\it{x}$ = 0, 0.1, 0.2, 0.5, and 1) with space group  $\textit{C2/m}$ (\#12).}

\begin{tabular}{ p{2.8 cm} p{2.5 cm}  p{2.5 cm}  p{2.5 cm} p{2.5 cm}  p{2.5 cm} p{2.5 cm} } \hline

Li$_4$CuTeO$_6$ & $a$ = 5.2692 (1) $\AA$ & $b$=  8.7887 (2)  $\AA$   & $c$ = 5.2298 (1)  $\AA$ & $\alpha$ = $\gamma$ =  90$^{\circ}$ & $\beta$ = 113.043 $^{\circ}$  & $\chi^2$ = 6.05 \\ 
&   & Cell Volume & =   222.872  $\AA^3$  &   &     & R$_{wp}$ = 0.064 \\
Atom  & wyck & \textit{x} & \textit{y} & \textit{z} & Occ. & B (\AA)  \\ \hline

  Te1   &  2a &  0.0000 &  0.0000 &  0.0000 &      1  & 0.0022 \\ 
  Li1  & 4h &   0.0000 & 0.1894  & 0.5000  &       1 & 0.0003 \\
  Li2/Cu2  & 2d &   0.5000 & 0.0000 & 0.5000  &  0.32/0.68 & 0.0016 \\
  Li3/Cu3  & 4g &  0.0000 & 0.3374 & 0.0000  &   0.84/0.16 & 0.0048 \\
  O1  & 4i &   0.2164  & 0.0000 & 0.7982  &  1 & 0.0063 \\
  O2 & 4j &  0.2308  & 0.1600 & 0.23812  &   1  & 0.0159 \\  \hline
  \\
  
Li$_4$Cu$_{0.9}$Ni$_{0.1}$TeO$_6$ & $a$ = 5.2675(1)   $\AA$  & $b$= 8.7862(3)  $\AA$ & $c$ = 5.2289 (2)  $\AA$ & $\alpha$ = $\gamma$ =   90$^{\circ}$ & $\beta$ = 113.04643 $^{\circ}$  & $\chi^2$ = 2.014 \\
&   & Cell Volume & =  222.691  $\AA^3$  &   &     &  R$_{wp}$ =  0.0686\\

   Te1   &  2a & 0.0000  &  0.0000 &  0.0000  &      1  & 0.00363 \\ 
  Li1  & 4h &   0.0000 & 0.1914 & 0.5000  &       1& 0.00223 \\
  Li2/Cu2  & 2d&   0.5000 & 0.0000 & 0.5000  &  0.32/0.68 & 0.00223 \\
  Li3  & 4g&  0.0000 & 0.3419 & 0.0000  &        0.85 & 0.00735 \\
  Cu3/Ni3 & 4g& 0.0000 &   0.3419 & 0.0000 &     0.11 /0.05  &  0.00735 \\
  O1  & 4i & 0.2080 & 0.0000  & 0.7689  &  1 & 0.00690 \\ 
  O2 & 4j& 0.2321 & 0.1548 & 0.2383  &   1  & 0.00690 \\ \hline  
  
   \\
  
Li$_4$Cu$_{0.8}$Ni$_{0.2}$TeO$_6$ & $a$ = 5.2586 (1)  $\AA$  &$b$ = 8.8030 (2)  $\AA$  & $c$ = 5.2237 (1)  $\AA$  & $\alpha$ = $\gamma$ =   90$^{\circ}$ & $\beta$ =  112.76 $^{\circ}$  & $\chi^2$ = 2.703 \\
&   & Cell Volume & = 222.991   $\AA^3$  &   &     &  R$_{wp}$ = 0.05 \\
   Te1   &  2a &  0.0000 &  0.0000 &  0.0000 &      1  & 0.0036 \\ 
  Li1    & 4h &   0.0000 & 0.1714 & 0.5000  &       1& 0.0010\\
  Li2/Cu2    & 2d &   0.5000 & 0.0000 & 0.5000  &  0.32/0.68 & 0.0057\\
  Li3  & 4g &  0.0000 & 0.3289  & 0.0000  &        0.84& 0.0099 \\
  Cu3/Ni3  & 4g &   &  & &    0.061/0.098  &  0.0090\\
  O1  & 4i & 0.2246 (8)    & 0.0000 & 0.7880 (3) &  1 & 0.0069\\ 
  O2 & 4j&  0.2346 (1) & 0.1561 (5) & 0.2455 (4)&   1  & 0.0077\\ \hline  
  
   \\
  
Li$_4$Cu$_{0.5}$Ni$_{0.5}$TeO$_6$  & $a$ =  5.2166  $\AA$ & $b$ = 8.8257  $\AA$  & $c$ = 5.1927 (3)  $\AA$ & $\alpha$ = $\gamma$ =   90$^{\circ}$ & $\beta$ = 111.902 $^{\circ}$ &$\chi^2$ = 3.213 \\
&   & Cell Volume & =  221.807  $\AA^3$  &   &     & R$_{wp}$ = 0.12 \\

  Te1   &  2a &  0.0000 &  0.0000 &  0.0000 &      1  & 0.0022 \\ 
  Li1  & 4h &   0.0000 & 0.1603 & 0.5000  &       1& 0.0003\\
  Li2/Cu2  & 2d&   0.5000 & 0.0000 & 0.5000  &  0.348/0.669& 0.0016 \\
  Li3/Ni3  & 4g &  0.0000 & 0.3156 & 0.0000  &        0.745/0.245& 0.0048 \\
  O1  & 4i &  0.2237560 & 0.0000 & 0.7688  &  1 & 0.0063\\ 
  O2 & 4j&  0.2302 & 0.1309 & 0.2353  &   1  & 0.0159\\ \hline  
   \\
  
Li$_4$NiTeO$_6$ & $a$ = 5.1579  $\AA$  & $b$ = 8.8871  $\AA$  & $c$ =  5.1397  $\AA$  & $\alpha$ = $\gamma$ =   90$^{\circ}$ & $\beta$ = 110.22 $^{\circ}$ & $\chi^2$ = 3.20\\
&   & Cell Volume & = 221.088   $\AA^3$  &   &     &  R$_{wp}$ = 0.06\\
  
 Te1   &  2a &  0.0000 &  0.0000&  0.0000 &      1  & 0.0036 \\ 
  Li1  & 4h &   0.0000 & 0.1578 & 0.5000  &       1& 0.003\\
  Li2  & 2d&   0.5000 & 0.0000 & 0.5000  &  1 & 0.0057 \\
  Li3/Ni3  & 4g&  0.0000 & 0.3352 & 0.0000  &   0.5/0.5 &0.0073 \\
  O1  & 4i &    0.2180 & 0.0000 & 0.7887  &  1 & 0.0069\\ 
  O2 & 4j&  0.2379 &  0.1546  & 0.2452   &   1  & 0.0070\\ \hline  \hline 
   \end{tabular}

 \label{table}
 \end{table*}

\begin{table*}[t]
\centering
\caption{ Comparison of the experimental and computed lattice constants, ${a}$, ${b}$, ${c}$, and cell volume for Li$_4$NiTeO$_6$ in the current study and literature data.} 
\begin{tabular}  { |p{3.5cm}|p{1.5cm}| p{1.5cm}|p{1.5cm}|  p{1.5cm}|p{2.5cm}|p{1.5cm}| p{1.5cm}|}	
 \hline
  & $a$ ($\AA$)   & $b$ ($\AA$) & $c$ ($\AA$) & $\beta$ ($^{\circ}$) & cell volume ($\AA^3$ ) & $\chi^2$ &    \\	\hline
Experiment & 5.1579 & 8.887   & 5.139  & 110.22   & 221.088 & 3.20 & this work \\ \hline	
Experiment-LeBail fit & 5.1603  & 8.8914   & 5.1426  & 110.209   & 221.429 &  &Ref.~\onlinecite{Uma} \\ \hline
Experiment & 5.1584  & 8.8806   & 5.1366  & 110.241   & 220.854 & 10.9 &Ref.~\onlinecite{Tarascon} \\ \hline
Experiment &  5.1568 &  8.8949  & 5.1388   &   110.107 & 221.350 &22.4 &  Ref.~\onlinecite{Bernd} \\ \hline
DFT, GGA +PBE  &  5.1216 & 8.9206   & 5.1397  & 110.283  & 220.208 & & Ref.~\onlinecite{Bao}\\ \hline
Std Dev (\%) & 0.016 & 0.015 & 0.002   & 0.065  & 0.489  &   & \\ \hline
\end{tabular} 
\label{table2}	
\end{table*}

In Li$_4$NiTeO$_6$, with the full Ni substitution ($x =$ 1), the crystal structure model used for Li$_4$CuTeO$_6$ does not yield a good fit. Previous studies found that due to a larger ionic radius difference between Li$^+$ and Ni$^{2+}$, Ni would preferably replace only a honeycomb ``4$g$" sites in Li$_4$CuTeO$_6$.\cite{Bernd} We utilize the structure model from Ref.~\onlinecite{Bernd} and find a significantly better fit as shown in Table I. The resulting structure is plotted in Fig.~\ref{unitcell} (b). The lattice parameters of Li$_4$NiTeO$_6$ from previous studies and their goodness of the fit are shown in Table ~\ref{table2}. Our lattice parameters are in excellent agreement with previous reports and we obtain the lowest $\chi^2 \sim$ 3.2, which may indicate a better phase quality of our sample. 

For the structural analysis of intermediate Ni doped ($x =$ 0.1, 0.2, and 0.5) compounds, if we allow Ni randomly goes into both 2$d$ and 4$g$ sites as in Li$_4$CuTeO$_6$, our Rietveld refinements give a high reliability parameter (goodness of fit), $\chi^2 \geq$ 20. However, if we restrict that Ni can only go into the 4$g$ site in the honeycomb layers and Cu can go into the 2$d$ site in the triangular layers similar to the model used in Li$_4$NiTeO$_6$,\cite{Bernd} we get very good fits with  $\chi^2$ = 2.014 $-$ 3.213. Intriguingly, this observation hints that Ni$^{2+}$ may prefer occupying the 4$g$ site in the honeycomb layers and share the site with Li$^+$ but not with Cu$^{2+}$. This indicates that the site-mixing at the 4$g$ site persists with increasing Ni concentration whereas the site-mixing at the 2$d$ site becomes less and finally disappears with full Ni substitution. 

\begin{figure}[h]
   \centering
    \includegraphics[width=1\linewidth]{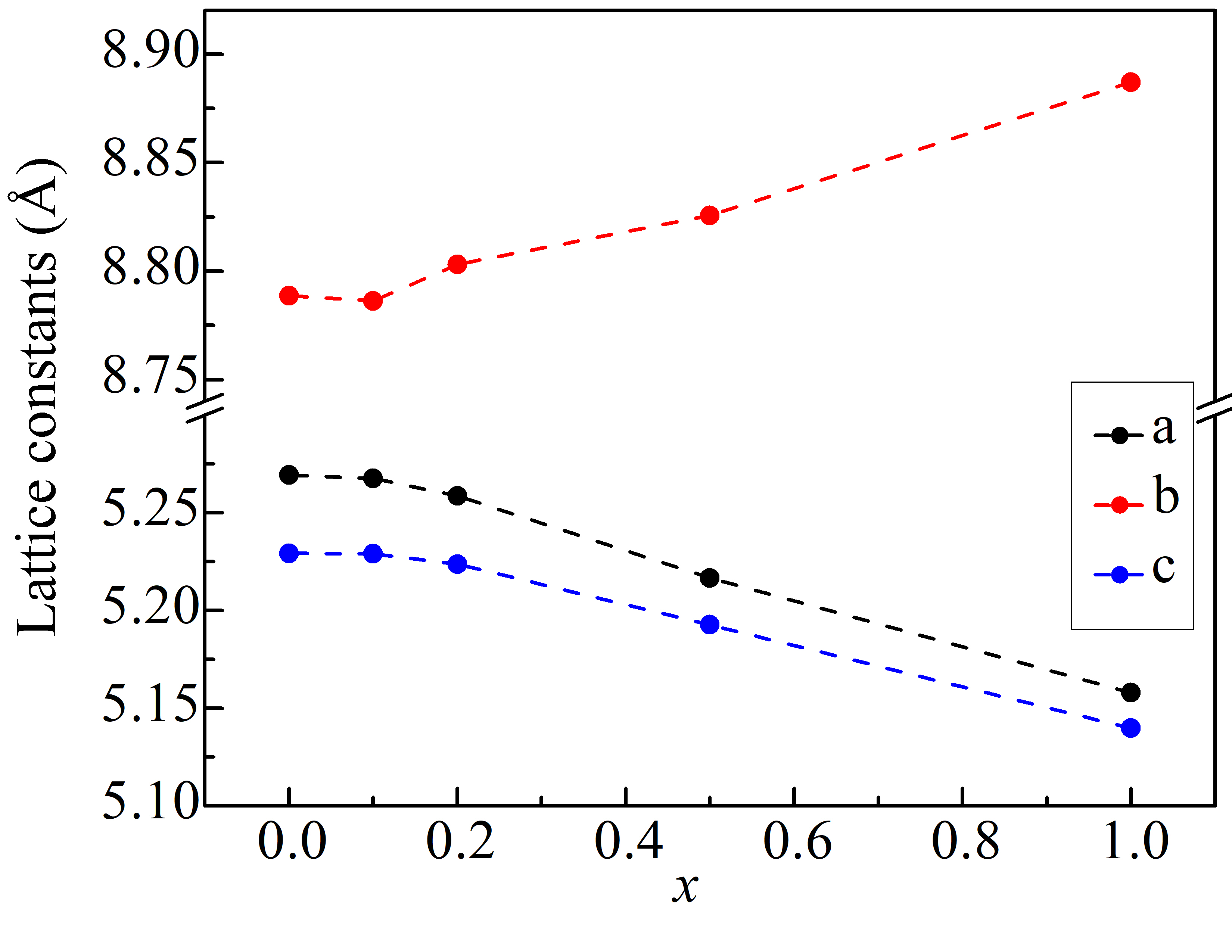} 
    \caption{Variation of lattice constants $a$, $b$, and $c$ as a function of Ni doping $x$ in  Li$_4$Cu$_{1-x}$Ni$_x$TeO$_6$.}
  \label{PXRD1}
\end{figure} 

Figure~\ref{PXRD1} presents the change of lattice parameters as a function of Ni doping levels. We find a slow increase (decrease) of the lattice parameters of $b$ ($a$ and $c$) below $x$ = 0.2. Then, the lattice parameter $b$ increases linearly above $x >$  0.2 while the lattice parameters $a$ and $c$ decrease linearly (Table ~\ref{table}). Such a reduction in the lattice parameters $a$ and $c$ is likely caused by the cation size difference between Ni$^{2+}$ (0.5569 $\AA$) and Cu$^{2+}$  (0.5773 $\AA$) ions. The full incorporation of Ni at the Cu site (4$g$) leads to a contraction in the unit cell volume by approximately 1\%.

\subsection{Magnetism}

\begin{figure}[h!]
   \centering
    \includegraphics[width=1\linewidth]{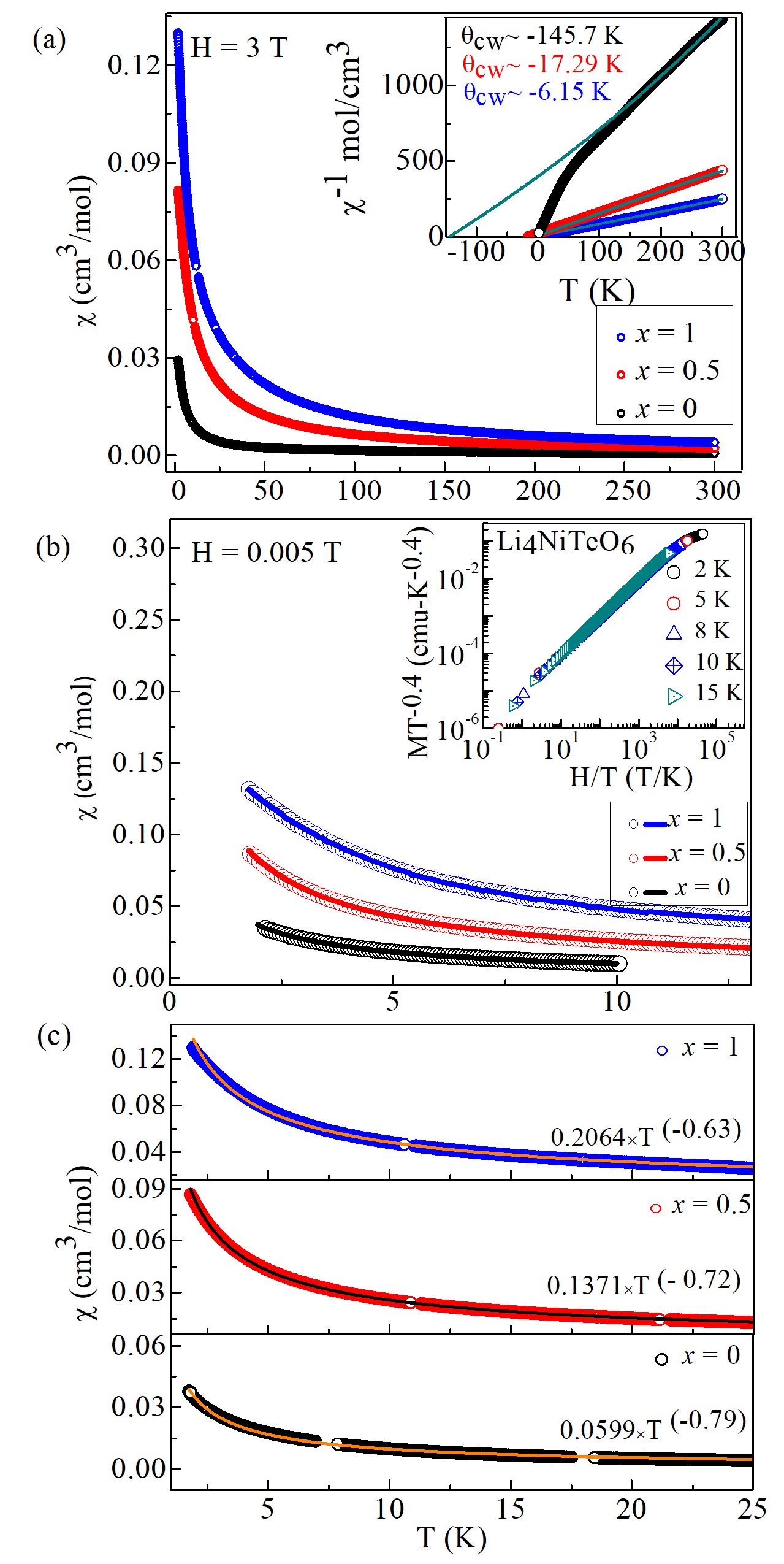}
   \caption{(a) Magnetic susceptibility $\chi$ versus $T$ for Li$_4$Cu$_{1-x}$Ni$_x$TeO$_6$ ($x$ = 0, 0.5, and 1) between $T$ = 1.8 and 300 K. The inset (i) shows $\chi^{-1}$ versus $T$. The solid gray curve through the data shows the Curie Weiss fit, $\chi$ = $\chi_0$ + $\frac{C}{(T-\theta)}$.  (b)  Zero-field-cooled (open circles) and field-cooled (lines) data measured in $H$ = 0.005 T for $x$ = 0, 0.5, and 1. Inset shows $T-H$ scaling of  $M(H)$ plotted on a log-log scale for $x$ = 1. (c) The solid lines in the $\chi$ versus $T$ data, represent a power-law fit to $\chi(T)$~$\sim~T^{-\alpha}$ for $T~\leq$~25 K.}
 \label{VSM}

\end{figure}

We show the temperature dependence of magnetic susceptibility ($\chi$) of Li$_4$Cu$_{1-x}$Ni$_x$TeO$_6$ ($x$ = 0, 0.5, and 1) measured under applied magnetic fields $H$~=~3~T in Fig.~\ref{VSM}\textcolor{blue}{(a)}. We measured magnetic susceptibility measurements at $H$ = 0.005 T (Fig.~\ref{VSM}\textcolor{blue}{(b)}), 1 T (not shown) and 3 T (Fig.~\ref{VSM}\textcolor{blue}{(a)}) and did not find any major changes in the magnetic behavior at low temperatures in all our measurements. Curie-Weiss-like behavior is clearly seen above $T~\geq$~100~K. We do not find any hint of magnetic order down to the base temperature of our measurement ($T~=~1.8$~K). We also confirm the absence of the spin-glass state down to $T~=~1.8$~K by measuring the zero-field-cooled (ZFC) and field-cooled (FC) magnetic susceptibility with $H$ = 0.005 T [Fig.~\ref{VSM}\textcolor{blue}{(b)}]. All samples do not show any kind of bifurcation in $\chi(T)$. This rules out the spin-glass transition in $x$ = 0, 0.5, and 1 samples.

The inverse susceptibility ($\chi^{-1}$) is plotted in the inset of Fig.~\ref{VSM}\textcolor{blue}{(a)}. We fit the data between $T$ = 150 K and 300  K using $\chi$ = $\chi_0$ + $\frac{C}{(T-\theta)}$. Our best fit results (the parameters $\chi_0$, $C$, and $\theta_\mathrm{CW}$) are shown in Table~\ref{table1}. For the undoped Li$_4$CuTeO$_6$ sample, we find $\chi_0$ = -1.867 $\times$ 10$^{-4}$ and $\theta_\mathrm{CW}$ $\approx$ -146  K which implies a strong AFM interaction in this compound. Our result is consistent with the previous report~\cite{Khuntia} although we get a slightly smaller value of the Weiss temperature (see Table~\ref{table1}). We get $C$ = 0.38 cm$^3$/mol, corresponding to an effective magnetic moment, $\mu_{eff}$ = 1.85(5) $\mu_B$ (assuming g = 2) which is slightly larger than the value for $\it{S}_{eff}$ = $1/2$.

While the temperature dependence of these materials looks similar at first glance, one can see an increase in the absolute magnitude of $\chi$(T) with Ni doping in Fig.~\ref{VSM}\textcolor{blue}{(a)}. Such increases in the Curie constants, $C$ can be due to the higher spin of Ni ($S$ = 1) and different magnetic interactions in the honeycomb layer in Li$_4$NiTeO$_6$. Additionally, the Weiss temperature $\theta_\mathrm{CW}$ decreases drastically from $\approx$ -145  K (for $x$~=~0) to -17  K (for $x$~=~0.5), and -6  K (for $x$~=~1), implying a strong suppression in the strength of the AFM interaction due to Ni substitution although the AFM interactions persist between the surviving local magnetic moments despite the change in the magnetic interactions from the combined triangular and honeycomb layers (Li$_4$CuTeO$_6$) to the honeycomb layer alone (Li$_4$NiTeO$_6$). For Li$_4$NiTeO$_6$ ($x$ =1), we find a negative value of $\chi_0$ = $-1.031~\times$~10$^{-4}$~cm$^3$/mol which is not consistent with Ref.~\onlinecite{Bernd} but matches well for the diamagnetic correction, \cite{Bain1, Bennett1} which we get 
 ($-1.02~\times$~10$^{-4}$~cm$^3$/mol).

\begin{table}[h]
\centering
\caption{Parameters obtained from fits to the magnetic susceptibility data by the Curie-Weiss expression $\chi$ = $\chi_0$ + $\frac{C}{(T-\theta)}$.  }
\begin{tabular}{| p {2.6 cm }| p {1.4 cm }|p{1.4 cm }| p {1.3 cm } |p {1.4cm }  |} \hline

  & $\chi_0$$\times$10$^{-4}$ cm$^3$/mol & $C$ cm$^3$/mol &  $\theta$  (K) & Reference | \\ \hline
Li$_4$CuTeO$_6$   &  -1.98  & 0.39  & -154 & Ref.~\onlinecite{Khuntia}     \\ \hline
Li$_4$CuTeO$_6$    &  -1.867  & 0.38  & -145.68  & this work  \\ \hline
Li$_4$Cu$_{0.5}$Ni$_{0.5}$TeO$_6$  & -0.3202 & 0.73 & - 17.29 & this work \\\hline
Li$_4$NiTeO$_6$   &  -1.031 & 1.26 & -6.15 & this work \\ \hline
Li$_4$NiTeO$_6$   &  1.48 & 1.069 & -11.4 & Ref.~\onlinecite{Bernd}  \\ \hline
 \end{tabular}
 \label{table1}
 \end{table}

Further, as $T~\rightarrow$~0~K, $\chi(T)$ exhibits a steep increase without a kink or saturation in all three materials, as shown in Fig.~\ref{VSM}\textcolor{blue}{(a)}. It is notable that below $T~\leq$~25 K, $\chi(T)$ obeys a power-law increase $\chi(T)$~$\sim~T^{-\alpha}$ (shown in Fig.~~\ref{VSM}\textcolor{blue}{(c))}, where ${\alpha}$~$\approx$~0.79, 0.72, and  0.63 for $x$ = 0, 0.5, and 1, respectively. In the absence of any sign of the Curie tail for $T~\leq$~25 K, such a proportionality $\chi(T)~\sim~T^{-\alpha}$ (0 $\leq$ ${\alpha}$ $\leq$ 1) indicates the presence of abundant low-energy excitations due to the power-law distribution of antiferromagnetic exchange energies in disorder driven random-singlet regime.~\cite{Lee, Kimchi, Kundu} 

In spin-1/2 quenched disordered quantum paramagnetic systems, the exponent $\alpha$ is observed to describe $MT^{-\alpha}$ versus $H/T$ for $T~\ll~H$.~\cite{Lee, Kimchi} We plot $MT^{-\alpha}$ versus $H/T$ for Li$_4$NiTeO$_6$ in the inset of Fig.~\ref{VSM}\textcolor{blue}{(b)}. We find $\alpha$ = 0.4 ($MT^{-0.4} \sim H/T$), which is not equal to but close to $\alpha = 0.63$ ($\chi(T)$~$\sim~T^{-\alpha}$). Such scaling parameters are argued to arise from the power-law probability distribution of antiferromagnetic effective exchange and were found to vary widely, 0 $\leq$ $\alpha$ $\leq$ 1 in different systems.~\cite{Lee, Kimchi} For example, $\alpha$ = 0.44 for LiZn$_2$Mo$_3$O$_8$,~\cite{Lee, Kimchi} $\alpha$ = 0.5 for ZnCu$_3$(OH)$_6$Cl$_2$,~\cite{Lee, Kimchi} $\alpha$ = 0.34 for Y$_2$CuTiO$_6$,~\cite{Kundu} $\alpha$ = 0.5 for H$_3$LiIr$_2$O$_6$,~\cite{Takagi} $\alpha$ = 0.72 for Cu$_2$IrO$_3$,~\cite{Choi} and $\alpha$ = 0.15 for Li$_4$CuTeO$_6$.~\cite{Khuntia} Despite the wide variety of $\alpha$ values, a hidden universal scaling parameter $q$, which describes the dependence to the spin-orbit coupling (SOC) and its spatial symmetry, is proposed by recent theory.~\cite{Lee, Kimchi} Without the SOC ($q$ = 0), it is proposed that the scaling parameters $\alpha$ = 0.5 in the $S$~=~1/2 quenched quantum magnets.  Our scaling parameters in Li$_4$NiTeO$_6$ are  very similar  to the proposed value for the spin-1/2 system, and one can think that two different $\alpha$ values might be due to the SOC or, more importantly, $S$ = 1 for Ni in Li$_4$NiTeO$_6$. A further theoretical study is required to understand the accidental similarity found here in the $S$ = 1 system. \\

\section{Summary and Discussion}

We investigated a series of polycrystalline Li$_4$Cu$_{1-x}$Ni$_x$TeO$_6$ ($x$ = 0, 0.1, 0.2, 0.5, and 1) compounds. We found no hint of magnetic ordering or freezing down to 1.8 K. Magnetic susceptibility measurements provide evidence that the Cu and Ni atoms carry effective spin, $S = 1/2$ and $S~=~1$, respectively. AFM exchange interactions persist in the entire Ni doping, evidenced by the change of the Curie-Weiss temperatures from $\theta_\mathrm{CW} = -145~$K [Li$_4$CuTeO$_6$ ($x~=~0)$] to $\theta_\mathrm{CW} = -6~$K [Li$_4$NiTeO$_6$ ($x~$=~1)]. The absence of any long-range magnetic order implies the randomness driven QSL-like state in Li$_4$Cu$_{1-x}$Ni$_x$TeO$_6$ ($x$ =0, 0.1, 0.2, 0.5, and 1) compounds, as suggested by Ref. \cite{Khuntia, Mourigal}. We also find a scaling nature of isothermal magnetisation  $M(H)$ ($T$ $\leq$ 15 K) for $x$ = 1 sample. Understanding the origin of this scaling behavior, especially in $x$ = 1 compound, requires further studies. 

Our detailed X-ray analysis shows that, in the parent Li$_4$CuTeO$_6$, Li and Cu are mixed randomly on both the 2$d$ and 4$g$ sites. With Ni doping (up to 50\%), we find that Ni ($S~=1$) substitutes for Cu ($S~= 1/2$) on the 4$g$ site while the Cu in the 2$d$ site remains almost unchanged. It implies that increasing Ni concentration does not impact the frustration caused by Cu ($S~=~1/2$) in the triangular layer. Due to the exchange of magnetic species in the honeycomb layer between Cu ($S~=~1/2$) and Ni ($S~=~1$), the origin of the AFM interactions changes from $S~=~1/2$ to $S~=~1$ and from the combined triangular and honeycomb layers (Li$_4$CuTeO$_6$) to the honeycomb layer alone (Li$_4$CuTeO$_6$). Therefore, it is likely that either the mechanism or the nature of quantum magnetic ground states in Li$_4$Cu$_{1-x}$Ni$_x$TeO$_6$ changes with increasing Ni concentrations. Not only the details of site-mixing in triangular and honeycomb layers but also the change of magnetic species play a crucial role in these materials.  Li$_4$Cu$_{1-x}$Ni$_x$TeO$_6$ are excellent materials to study systematically how this exotic magnetic ground state evolves between different spin states, between different geometrical frustrations, and in a combination of spin states and geometrical frustrations.

\section{ Acknowledgment}
This work is supported by the University of Wisconsin-Milwaukee. Work at the University of Arizona is supported by the University of Arizona startup fund.

\section{DATA AVAILABILITY}
The data supporting this study's findings are available within the article.

\end{document}